\documentclass[prl,twocolumn,aps,showpacs,nofootinbib,floatfix,nobibnotes]{revtex4-2}
\usepackage{amsmath, bm}
\usepackage{amssymb}
\usepackage{graphicx}
\usepackage{verbatim}
\usepackage[colorlinks=true,linkcolor=blue,citecolor=blue,urlcolor=blue]{hyperref}
\usepackage{txfonts}
\usepackage{dsfont}
\usepackage{enumitem}
\usepackage[utf8]{inputenc}

\AtBeginDocument{%
    \newwrite\bibnotes
    \def\bibnotesext{Notes.bib}
    \immediate\openout\bibnotes=\jobname\bibnotesext
    \immediate\write\bibnotes{@CONTROL{REVTEX41Control,eprint=""}}
    \immediate\write\bibnotes{@CONTROL{%
    apsrev41Control,author="08",editor="1",pages="1",title="0",year="1"}}
     \if@filesw
     \immediate\write\@auxout{\string\citation{apsrev41Control}}%
    \fi
}%

\usepackage{titlesec}
\titleformat{\section}
  {\centering\normalfont\large\bfseries}{\thesection.}{1em}{}
\titleformat{\subsection}
  {\centering\normalfont\normalsize\bfseries}{\thesubsection.}{1em}{}
\titleformat{\subsubsection}
  {\centering\normalfont\normalsize\itshape}{\thesubsubsection.}{1em}{}

\begin{document}

\title{Quantum metrology of two-photon absorption}
\author{Carlos S\'anchez~Mu{\~n}oz$^1$}
\email{carlossmwolff@gmail.com}
\author{Gaetano Frascella$^{2,3}$}
\author{Frank Schlawin$^{4,5}$}
\email{frank.schlawin@mpsd.mpg.de}
\affiliation{$^1$ Departamento de F\'isica Te\'orica de la Materia Condensada and Condensed Matter Physics Center (IFIMAC), Universidad Aut\'onoma de Madrid, Madrid, Spain}
\affiliation{$^2$ Max-Planck Institute for the Science of Light, Staudtstr. 2, Erlangen D-91058, Germany}
\affiliation{$^3$ University of Erlangen-Nuremberg, Staudtstr. 7/B2, Erlangen D-91058, Germany}
\affiliation{$^4$ Max Planck Institute for the Structure and Dynamics of Matter, Luruper Chaussee 149, 22761 Hamburg, Germany }
\affiliation{$^5$ The Hamburg Centre for Ultrafast Imaging, Luruper Chaussee 149, Hamburg D-22761, Germany}

\date{\today}

\begin{abstract}

Two-photon absorption (TPA) is of fundamental importance in super-resolution imaging and spectroscopy. 
Its nonlinear character allows for the prospect of using quantum resources, such as entanglement, to improve measurement precision or to gain new information on, e.g., ultrafast molecular dynamics.
Here, we establish the metrological properties of nonclassical squeezed light sources for precision measurements of TPA cross sections. 
We find that there is no fundamental limit for the precision achievable with squeezed states in the limit of very small cross sections. 
Considering the most relevant measurement strategies---namely photon counting and quadrature measurements---we determine the quantum advantage provided by squeezed states as compared to coherent states. We find that squeezed states outperform the precision achievable by coherent states when performing quadrature measurements, which provide improved scaling of the Fisher information with respect to the mean photon number $\sim n^4$. Due to the interplay of the incoherent nature and the nonlinearity of the TPA process, unusual scaling can also be obtained with coherent states, which feature a $\sim n^3$ scaling in both quadrature and photon-counting measurements.  

\end{abstract}

\maketitle

{\it Introduction}--Two-photon absorption (TPA), the simultaneous absorption of two quanta of light by a quantum system, was first described theoretically by Maria Goeppert-Mayer in 1931~\cite{Goeppert-Mayer}, and first observed experimentally only one year after Maiman's development of the laser~\cite{Kaiser61}. 
It has since become a crucial tool in spectroscopy and microscopy, where the nonlinear nature of TPA enables enhancing the resolution beyond the single-photon diffraction limit~\cite{So00}. 
TPA also forms one of the main fields of interest for the development of future quantum-enhanced photonic technologies, and, in particular, it is considered a possible application of entangled photon sources in imaging applications. It was recognised already in the later 1980's that the absorption probability of entangled photon pairs scales linearly with the light field intensity~\cite{Klyshko1982, Gea89, Javainen90, Georgiades95, Georgiades97}. This could enable nonlinear spectroscopy and microscopy at low photon fluxes, which will be beneficial in photosensitive samples and reduce phototoxicity in live organisms~\cite{Taylor2016, Dorfman16, JPhysB17, AccChemRes, Gilaberte2019, Mukamel2020, Szoke2020}. 

Quantum-enhanced absorption measurements have received renewed attention recently~\cite{Dinani2016,Whittaker2017,Sabines2017,Losero2018,Birchall20, Li20, Li20b, Shi20, Okamoto2020} with the development of new quantum light sources and an increased interest in sensing technologies~\cite{Chekhova16}, as well as the demonstration of ``sensing with undetected photons"~\cite{Lemos2014, Kalashnikov2016, Lindner2020, Kutas2020, Lindner2021}. 
Interest in this problem dates back to 2007, where the optimal estimation of single photon losses was first considered~\cite{Monras07, Adesso09}. 
The quantum Fisher information (QFI) of absorption measurements was evaluated. 
Cramer-Rao bounds for dissipative processes were determined~\cite{Escher2011, Alipour14}. Precision limits of phase estimation in the presence of interactions were established in~\cite{Boixo2008, Anisimov10}, which noted that interactions can enable so-called ``super-Heisenberg scaling" with the photon number $n$ in the sense that the optimal scaling of linear phase estimation precision ($\sim n^{-1}$) can be surpassed. 
These studies concern linear spectroscopy, i.e. the absorption of single photons, or the combination of classical lasers with quantum light sources in two-photon Raman transitions~\cite{deAndrade2020, Prajapati2020}. 
First theoretical works also started to investigate the role of quantum correlations in TPA of entangled photon pairs \cite{NJP17, Oka2018}, or of photon statistics in coherent control~\cite{Csehi19}. 
However, these first studies fall short of providing a comprehensive understanding of the role of photon statistics and correlations in nonlinear spectroscopy and in particular in TPA. Despite its importance for both existing and future imaging technologies, no quantum metrological bounds for TPA measurements exist to date. 

\begin{figure}[t]
\centering
\includegraphics[width=0.49\textwidth]{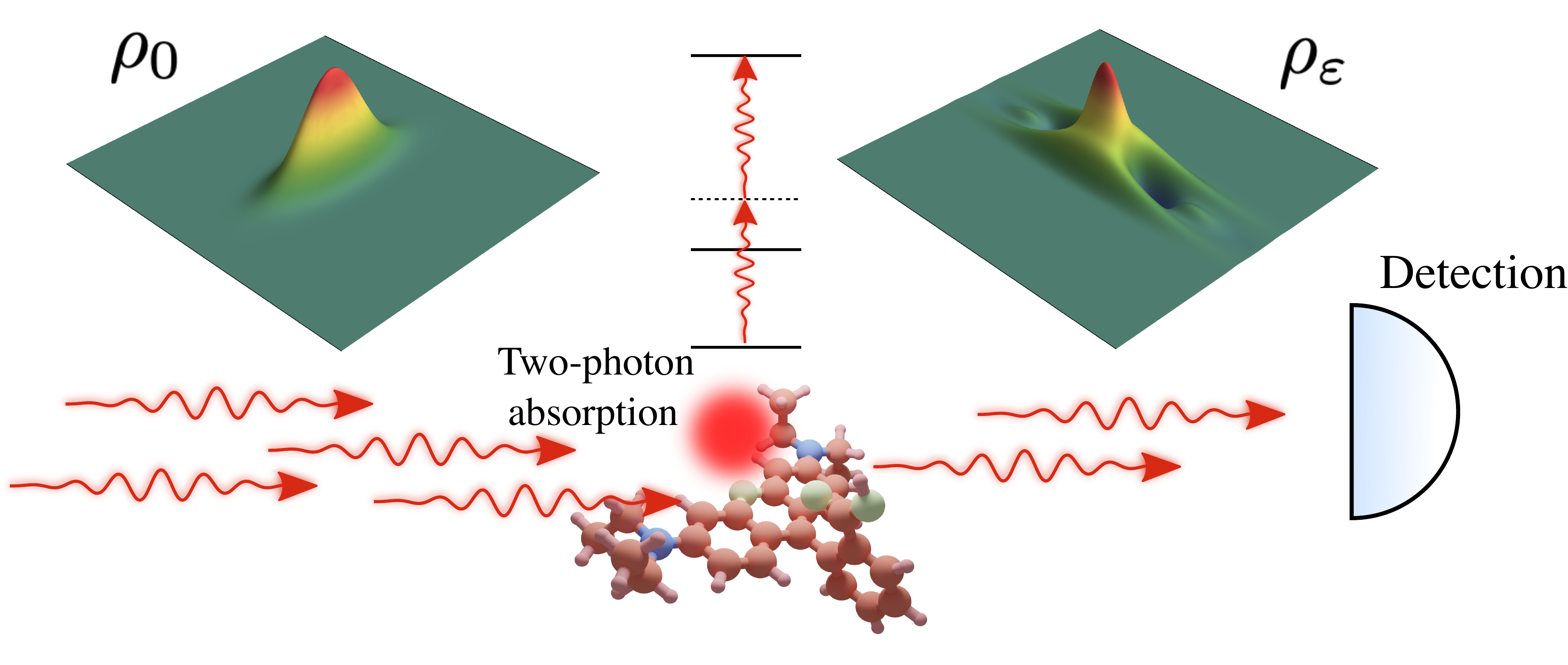}
\caption{(Color online) 
Sketch of the investigated setup: An initial quantum state of light $\rho_0$ (here a squeezed state) evolves under two-photon absorption (TPA) into $\rho_\varepsilon$ before being measured. 
}
\label{fig.setup}
\end{figure}

In this Paper, we derive quantum metrological bounds on the determination of two-photon absorption cross sections of narrowband light fields, and establish the metrological advantage provided by illumination with squeezed states of light. 
As pointed out already in a series of older publications~\cite{Gilles94, Guerra97, Enaki98, Jacobs06}, TPA losses can create a certain amount of non-Gaussianity in the transmitted field (see Fig.~\ref{fig.setup}). The evolution creating nonclassicality provides an additional layer of potential complexity, and stands in contrast to conventional phase estimation problems of coherent dynamics or losses, where the nonclassicality of the injected quantum state is generically either unaffected or reduced by the evolution. 
Therefore, TPA measurements constitute a fascinating metrological problem, which has not been considered to date. 

We find that, under TPA, the QFI scales with the mean photon number $n$ of the input light state as $\propto n^3$, when coherent states are employed. 
This scaling of the QFI surpasses the Heisenberg limit of $\propto n^2$, and is enabled by the non-unitary character of the evolution under TPA and does not occur in a coherent second-harmonic generation (SHG) process. 
One can saturate this QFI by both photon number and quadrature measurements. Moreover, we show this scaling can be improved even further using squeezed states, where the QFI diverges in the limit of very weak TPA losses and homodyne measurements of the squeezed quadrature provide a $\propto n^4$ scaling of the corresponding Fisher information. 

{\it Two-photon absorption}--We are interested in the situation sketched in Fig.~\ref{fig.setup}. A quantum state of light is transmitted through a TPA sample, then a measurement is carried out. We assume that there is no resonant intermediate state in the sample, such that single photon losses can be neglected. We also assume a narrowband light field which can be described by a single bosonic mode.
To describe this setup we integrate out the absorbing material using the normal methods of open quantum systems to obtain a Lindblad equation in the rotating frame with respect to the optical field Hamiltonian, which reads~\cite{Simaan75, Simaan75b, Simaan78, Gilles93}, 
\begin{align}
\frac{d}{dt} \rho &= \gamma_{\mathrm{TPA}} \mathcal{L} \rho = \frac{\gamma_{\mathrm{TPA}}}{2} \left( 2 L \rho L^{\dagger} - L^{\dagger} L \rho - \rho L^{\dagger} L \right), \label{eq.Lindblad}
\end{align}
where the Lindblad operator is given by a two-photon loss operator $L = a^2 / \sqrt{2}$. Our objective will be to measure the absorbance $\varepsilon \equiv \gamma_{\mathrm{TPA}} t$, where $t$ is the propagation time through the sample. This allows to determine the absorption cross section through $\sigma_Q = \varepsilon / (n \ell)$, where $n$ is the TPA sample density and $\ell$ the length of the sample medium in the propagation direction of the light. 

Our work focuses mainly on the metrological advantage of squeezed states of light, in which one quadrature features fluctuations below the shot-noise limit, at the expense of increased fluctuations in the opposite quadrature.
The time evolution of a squeezed state of light undergoing TPA losses is sketched in Fig.~\ref{fig.setup}, where the initial Wigner function is shown on the left, and the output Wigner function after TPA losses with $\varepsilon = 0.1$ on the right. 
When the squeezed state evolves according to the master equation~(\ref{eq.Lindblad}), the squeezed vacuum fluctuations will increase, whereas the anti-squeezed quadrature fluctuations are reduced. Crucially, however, this happens in a nonclassical way: 
notable negative areas in phase space develop on the sides of the initially squeezed quadrature direction, signifying non-Gaussianity of the output state. This strongly contrasts with the evolution under single photon losses, or the action of a coherent squeezing operation, which do not create negative values in the Wigner function. As a consequence, the measurement of TPA losses represents a fascinating quantum metrological problem of a fundamentally distinct, dissipative process.

{\it Fundamental sensitivity limits}--We first turn to the ultimate precision limit for the detection of TPA detection. The sensitivity $\Delta\varepsilon$ with which one can estimate $\varepsilon$ 
is given by the quantum Cramer-Rao bound \cite{Braunstein94, Paris09, Haase16},
\begin{equation}
\Delta\varepsilon^2 \geq 1 / \mathcal{F}_\rho, \label{eq.CramerRao}
\end{equation} 
where $\mathcal{F}_{\rho}$ is the quantum Fisher information (QFI) associated with the estimation of $\varepsilon$~\cite{Paris09}. 
This quantity can be obtained for arbitrary states $\rho_{\varepsilon}$ (denoting the state of light encoding the value $\varepsilon$) by constructing the symmetric logarithmic derivative (SLD) $L_\varepsilon$, 
which is defined by the equation
\begin{equation}
\frac{d\rho_{\varepsilon}}{d\varepsilon}=\frac{1}{2}\left(L_{\varepsilon}\rho_{\varepsilon}+\rho_{\varepsilon}L_{\varepsilon}\right),\label{eq:SLD}
\end{equation}
to yield the QFI $\mathcal{F}_{\rho} = \text{Tr} [ L_\varepsilon^2 \rho_{\varepsilon} ]$.
One can diagonalise $\rho_{\varepsilon} = \sum_{k} \lambda_k |k\rangle \langle k|$ to find
\begin{equation}
L_{\varepsilon} = 2 \sum_{k,l} \frac{ \langle l | ( \mathcal{L} \rho) |k\rangle }{ \lambda_k + \lambda_l } |l \rangle \langle k |. \label{eq:Lepsilon}
\end{equation}
For coherent states of light with complex amplitude $\alpha$ and photon number $n_\alpha = |\alpha|^2$, we can carry out this construction analytically in the limit of small TPA absorption, to obtain
\begin{equation}
\mathcal{F}_{\rho_{\mathrm{coh}}} (\varepsilon = 0) = n_\alpha^3 + \frac{n_\alpha^2}{2}, \label{eq.QFI_coh}
\end{equation}
which, notably, displays a scaling $\propto n^3$. This scaling is consistent with earlier results concerning phase estimation in the presence of two-body interactions~\cite{Boixo2007, Boixo2008, Boixo2008b}, according to which a scaling of the sensitivity $\Delta\varepsilon^2 \sim n^{-3}$ in the absence of entanglement is expected. 
However, this result does not generalise straightforwardly to measurements with entangled probes, where two-body interactions are expected to give rise to a $n^{-4}$-scaling of the achievable sensitivity. 
For a non-classical input state, such as the squeezed vacuum, we find that in general the QFI is not a very useful boundary for practical purposes:
While the QFI of a coherent state approaches a finite value for $\varepsilon \rightarrow 0$, we find that the QFI for a squeezed vacuum state, $S (\zeta) \vert0\rangle$, with $S (\zeta) = \exp (\zeta^\ast a^2 - \zeta a^{\dagger 2})$  and squeezing parameter $\zeta = r e^{i \varphi}$ diverges.  
Hence, there is \textit{no fundamental lower bound} on the precision with which small TPA losses can be detected. As we demonstrate in the SI, the divergence of the QFI can be traced back to the generation of a finite weight for non-Gaussian squeezed Fock state populations $\sim S (\zeta) |2\rangle\langle2|S^\dagger (\zeta)$ due to transient evolution with respect to Eq.~(\ref{eq.Lindblad}) already to linear order in $\varepsilon$. Therefore, a projective measurement on this non-Gaussian state is optimal and can determine very small absorbances $\varepsilon \ll 1$ without any fundamental lower error within the range of validity of the Markovian master equation~(\ref{eq.Lindblad}). Any fundamental error source would have to stem from non-Markovian effects.
A similar effect does not occur for coherent probe states, and consequently no divergence takes place in this case.
This remarkable result is a direct consequence of the incoherent nature of the TPA process and cannot be found, e.g., in coherent nonlinear processes such as second-harmonic generation (SHG). To see this, consider the SHG Hamiltonian of the form $H_{\mathrm{SHG}} \sim a^2 b^\dagger + h.c.$, where $b^\dagger$ is the photon creation operator of the SHG field. For a pure input state, the corresponding QFI is proportional to the variance of the Hamiltonian, $\mathcal{F}^{(\mathrm{SHG})}_{\rho} = 4( \langle H_{\mathrm{SHG}}^2 \rangle_\rho - \langle H_{\mathrm{SHG}} \rangle^2_\rho)$. Without loss of generality, we assume that the SHG field is in the vacuum, this variance evaluates for a squeezed vacuum to $\mathcal{F}^{(\mathrm{SHG})}_{\rho_{squ}} \sim  n^2$, which does not show super-Heisenberg scaling, nor can it ever diverge. 
Likewise, the QFI for coherent input evaluates to $\mathcal{F}^{(\mathrm{SHG})}_{\rho_{\mathrm{coh}}} \sim n^2$ and does not show super-Heisenberg scaling either. 

The positive operator-valued measurement (POVM) that saturates the precision limit established by QFI can be forbiddingly complicated to express or implement in practice (such as the projection on the state $S (\zeta) |2\rangle\langle2|S^\dagger (\zeta)$ in our case). 
Hence, to assess the metrological advantage of non-classical inputs for the measurement of TPA losses, it is instead necessary to consider particular measurement scenarios to derive practical bounds on the TPA precision for measuring a general operator $\mathcal{O}$,
\begin{align}
\Delta\varepsilon_{\mathcal{O} }^2 &= \frac{ \text{ Var } (\mathcal{O}) }{ \left| \frac{\partial \langle \mathcal{O} \rangle}{ \partial \varepsilon } \right|^2}. \label{eq.sensitivity}
\end{align}

{\it Photon number measurements}--The experimentally most relevant situation is a photon number measurement, where TPA losses are detected through a change in the transmitted photon number distribution. 
First, we thus consider the change of the mean transmitted photon number $ \langle \hat{n} \rangle = \langle a^\dagger a \rangle$. For a squeezed vacuum state, we find, to leading order in $\epsilon$, $\langle  \hat{n} \rangle_{squ} = n_r - \epsilon n_r (1 + 3 n_r)$, where $n_r = \sinh^2 (r)$. The variance is given by $\mathrm{Var}_{\mathrm{squ}} (\hat{n}) = \sinh^2 (r) [1 + \cosh (2r)]$, such that the sensitivity for the mean photon number reads
\begin{align}
\Delta \varepsilon^{2 (\mathrm{squ})}_{\hat{n} } &= \frac{2}{ n_r } \frac{1 + n_r }{ (1 + 3 n_r)^2 }. \label{eq.DeltaEpsilon_phN}
\end{align}
An identical calculation for a coherent state with complex amplitude $\alpha$ yields
\begin{align}
\Delta \varepsilon^{2 (\mathrm{coh})}_{\hat{n} } &= \frac{1}{ n_{\alpha}^3 }.
\end{align}
Hence, photon number measurements of coherent states already saturate the scaling of the corresponding QFI~(\ref{eq.QFI_coh}).
Perhaps counterintuitively, the sensitivity scaling of squeezed light for photon counting measurements $\sim n^{-2}$ is \textit{worse} than that of coherent light. Instead, it is the same as the scaling we obtained above for the QFI of SHG measurements with squeezed vacuum.

\begin{figure}[t]
\centering
\includegraphics[width=0.48\textwidth]{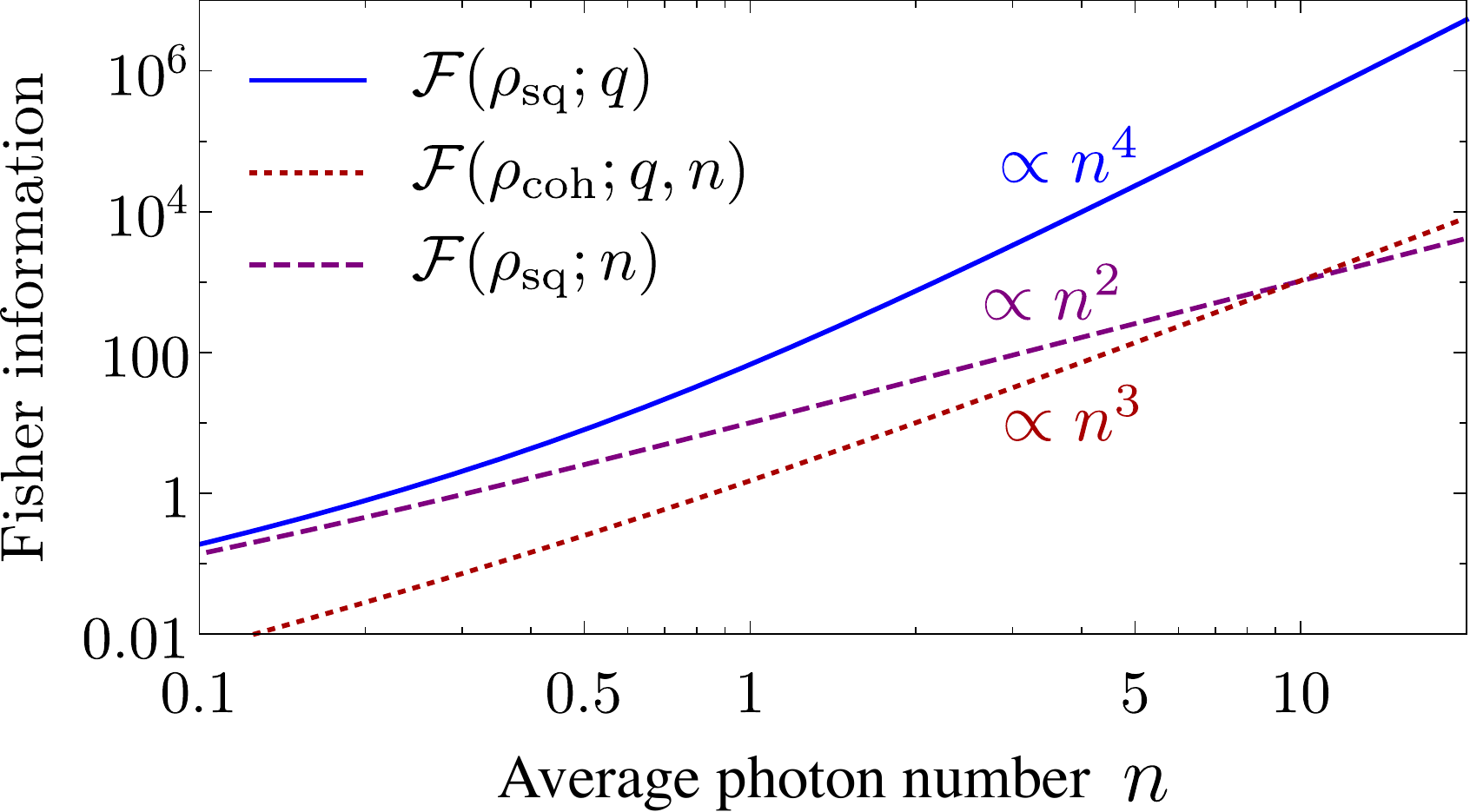}
\caption{(Color online) 
Classical Fisher information (CFI) vs. the mean photon number for quadrature and photon number measurements with squeezed input states, giving rise to quartic $\propto n^4$ and quadratic $\propto n^2$ scaling, respectively, as well as coherent states, which feature cubic scaling $\propto n^3$. These scaling are obtained in the limit of very weak TPA losses, i.e. $\varepsilon \rightarrow 0$.
}
\label{fig.phot-num}
\end{figure}

These different scaling behaviours cannot be improved by a full knowledge of the photon-number distribution. This can be seen by analysing the corresponding classical Fisher information (CFI),
\begin{align}
\mathcal{F}_C (\rho_\varepsilon, \hat{n}) &= \sum_n P_n (\varepsilon) \left( \frac{d}{d\varepsilon} \ln P_n (\varepsilon) \right)^2, 
\label{eq.F_C}
\end{align}
which bounds the inverse of the sensitivity for a given POVM. Here, the POVM is given by $\{|n\rangle \langle n| \}$, with $|n\rangle$ the $n$-photon Fock state, and the corresponding probabilities are given by $P_n (\varepsilon) = \langle n \vert e^{\varepsilon \mathcal L}\rho_0 \vert n \rangle$, describing the probability of detecting $n$ photons in transmission. Using Eq.~(\ref{eq.Lindblad}), we find that $dP_n / d\varepsilon \propto  (n+1) (n+2) P_{n+2}  - n (n-1) P_{n}$. As a consequence, narrow photon number distributions appear beneficial for detecting TPA losses, as they create large ``gradients" $P_{n+2}-P_n$ that enhance the Fisher information. This is why coherent light can outperform squeezed vacuum states in photon number measurements. 

This can be seen in Fig.~\ref{fig.phot-num}, where the corresponding Fisher information for squeezed and coherent states are plotted vs. their respective photon number expectation values. 
We find that the classical Fisher information $\mathcal F(\rho,\hat n)$ for squeezed light scales quadratically, $\propto  n^2$, while for coherent light it coincides with the QFI in Eq.~(\ref{eq.QFI_coh}), i.e. $\propto n^3$. 
As a consequence, coherent light outperforms squeezed light for photon-counting measurements at photon numbers $ n  \geq 10$ at small $\varepsilon$. With increasing absorbance, this crossover decreases to smaller photon numbers.  
The effect is not related to the crossover from linear to quadratic scaling of the squeezed light TPA absorption rate~\cite{Georgiades95}, which already takes place at $\langle n \rangle \gtrsim 1$. 
It is rather a consequence of the fact that Eq.~(\ref{eq.F_C}) favours narrow photon number distributions. 
Hence, it appears that from a quantum metrological perspective the use of squeezed states for TPA detection with photon number measurements only offers an advantage for small intensities. However, they do offer a significant advantage for quadrature measurements, as we show next. 

{\it Quadrature measurements}--We now turn to the measurement of field position $q = (a + a^\dagger)/\sqrt{2}$ and momentum quadrature $p = (a - a^\dagger)/(\sqrt{2}i)$. For a squeezed vacuum, the expectation value of either quadrature is zero, $\langle p \rangle = \langle q \rangle = 0$, and TPA will not shift this expectation value as it cannot create coherence.
Nevertheless, the analysis of the probability distributions associated to measurements of $p$ and $q$ contains vital information: using the Wigner function representation of the light fields~\cite{Brune1992, Garraway1992}, we find analytical expressions of the CFI at $\varepsilon = 0$, which for large photon numbers scale as
\begin{equation}
\mathcal{F}_C (\rho_{\mathrm{squ}}, q) \sim 32 n_r^4, \label{eq.F_C(X)}
\end{equation}
for the squeezed quadrature, and 
$\mathcal{F}_C (\rho_{\mathrm{squ}}, p) \sim 21 n_r^2/2$ for the anti-squeezed field quadrature. 
The full expressions are given in the SI. The precision for measurements with coherent states again saturates Eq.~(\ref{eq.QFI_coh}), $\mathcal{F}_C (\rho_{\mathrm{coh}}, q) = n_{\alpha}^3 + n^2_{\alpha} / 2$ for the displaced quadrature and  $\mathcal{F}_C (\rho_{\mathrm{coh}}, p) = n^2_{\alpha} / 2$ for the orthogonal quadrature, i.e. there is no improvement compared to photon number measurements discussed before. Thus, quadrature measurements of the squeezed quadrature can outperform coherent light and, in principle, achieve better sensitivity scaling than either photon number or quadrature measurements of coherent states.

\begin{figure}[t]
\centering
\includegraphics[width=0.48\textwidth]{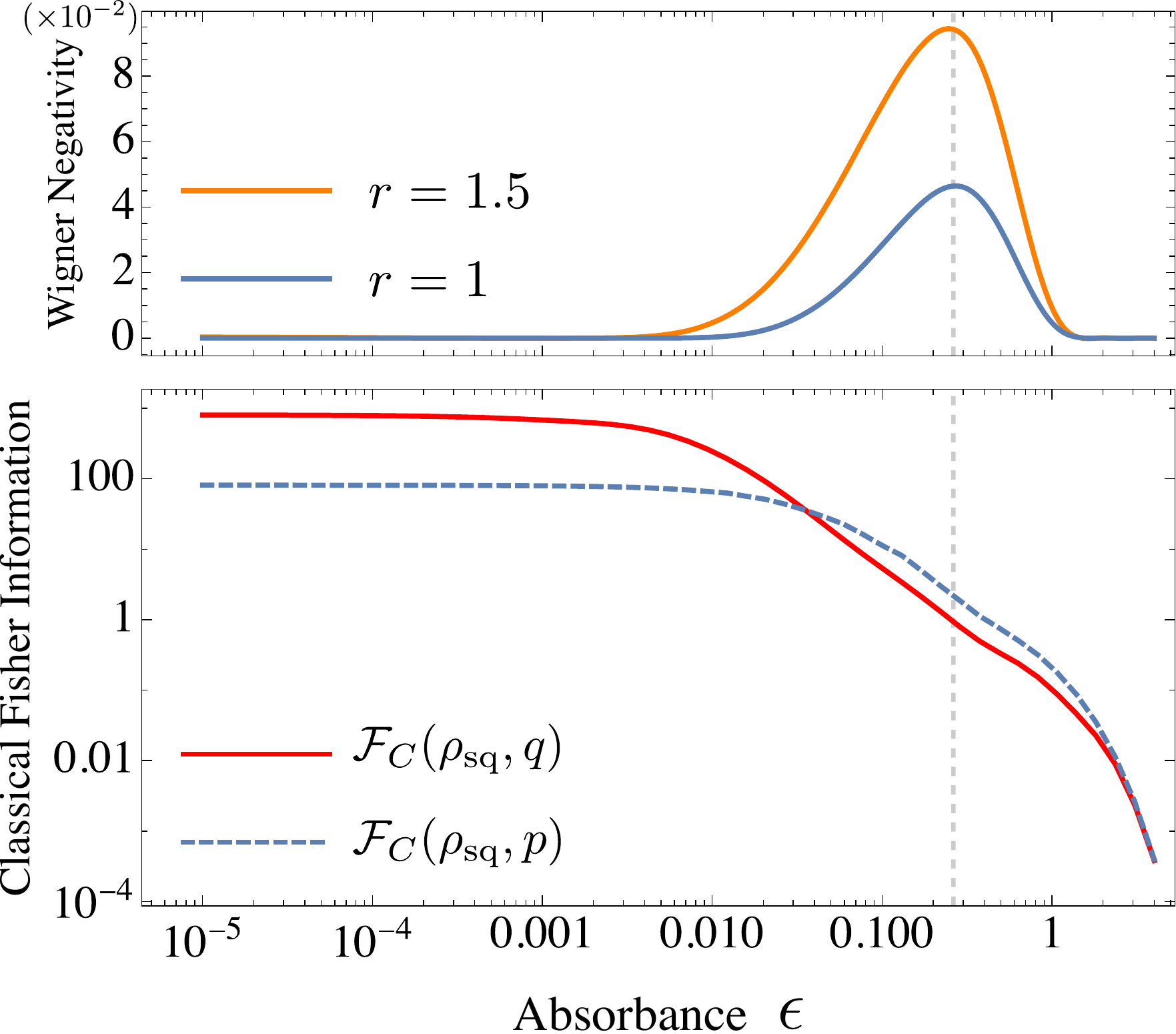}
\caption{(Color online) 
Top panel: Negativity of the Wigner function for an initial squeezed state with squeezing parameter $r = 1$ or $\langle n \rangle \simeq 1.4$ (blue), and $r = 1.5$ or $\langle n \rangle \simeq 4.5$ (orange). 
Bottom panel: Classical Fisher Information (CFI) vs. the TPA absorbance $\varepsilon = \gamma_{\mathrm{TPA}} t$ for initial squeezed state with squeezing parameter $r = 1$ for measuring the squeezed $q$-quadrature (red, solid) and the anti-squeezed $p$-quadrature (blue, dashed).
}
\label{fig.quadrature}
\end{figure}

In Fig.~\ref{fig.quadrature}, we investigate how this behaviour changes with the absorbance $\varepsilon$. 
It shows the evolution of the two quadratures as a function of $\varepsilon$. 
Measurements of the squeezed quadratures are superior only for small $\varepsilon \lesssim 10^{-2}$, i.e. when less than $1- \exp (- 10^{-2}) \simeq 0.1 \%$ of the signal have been absorbed. 
At larger $\varepsilon$, the CFI of the anti-squeezed quadrature becomes larger. Incidentally, as can be seen in the top panel of Fig.~\ref{fig.quadrature}, this crossover coincides with the emergence of negativity in the Wigner function, i.e. non-Gaussianity, of the quantum state of light. At even larger absorbances $\varepsilon \gtrsim 1$, the Fisher information of both quadratures merge, as the negativity disappears again and the quantum state of light is reduced to the vacuum state. 

\begin{figure}[t]
\centering
\includegraphics[width=0.48\textwidth]{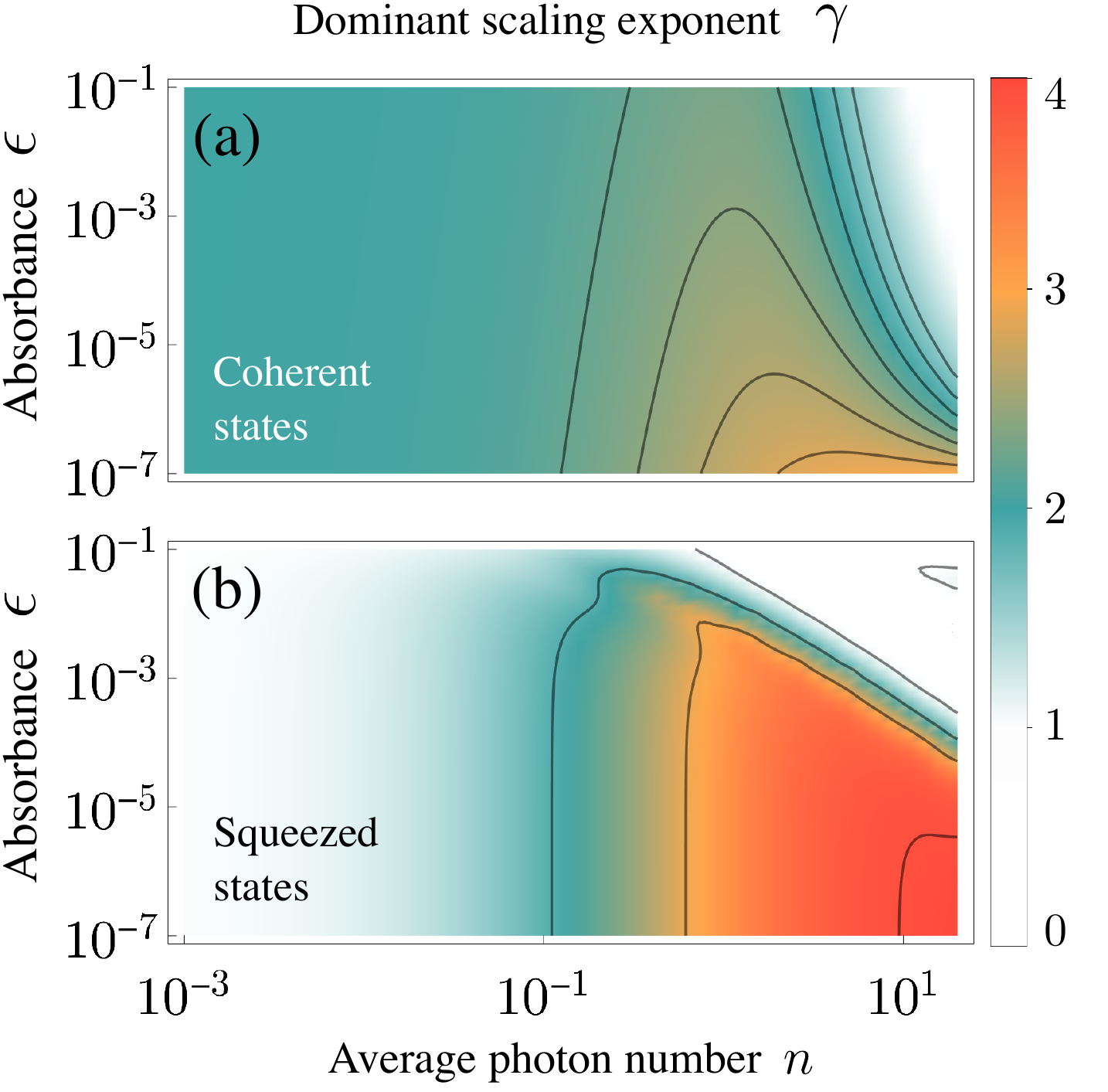}
\caption{(Color online) 
(a) Dominant scaling exponent, Eq.~(\ref{eq.exponent}), of the classical Fisher information~(\ref{eq.F_C}) for quadrature measurements of the squeezed quadrature is plotted vs. the photon number $n_0$ and absorbance $\varepsilon$ of a squeezed vacuum state.
(b) The same as in (a) for a coherent input state. 
(c) Classical Fisher information~(\ref{eq.F_C}) for quadrature measurements is plotted vs. the photon number $n_0$ and absorbance $\varepsilon$ of a squeezed vacuum state.
(d) The same as in (c) for a coherent input state. 
}
\label{fig.phase-space}
\end{figure}

At finite absorbance $\varepsilon$, the $n^4$-scaling in Eq.~(\ref{eq.F_C(X)}) is eroded concomitantly with the emergence of non-Gaussianity. We investigate this in Fig.~\ref{fig.phase-space} where we extract the dominant scaling exponent of the CFI using the derivative
\begin{equation}
\gamma = \frac{ \partial \log F (q) }{\partial \log n}. \label{eq.exponent}
\end{equation}
Naturally, to observe the $n^4$-scaling of squeezed vacuum or the $n^3$-scaling of coherent states numerically, we require a substantial photon number $n\sim 10$, as otherwise lower orders of the polynomial expansion remain dominant. These optimal scalings are eroded very quickly at large photon numbers (faster in the case of coherent states), while there is an intermediate regime at $n\sim 1$, where super-Heisenberg scaling can be sustained up to $\varepsilon = 10^{-2}$. As we show in the SI, where we plot the absolute value of the CFI, this only applies to the scaling, the absolute value of the CFI never decreases with increasing photon numbers.

{\it Conclusions}--We have examined precision bounds on the measurement of two-photon absorption cross sections. We focused in particular on the possible use of squeezed light for quantum-enhanced measurements. 
Remarkably, we found that there is no fundamental lower bound on the achievable precision of TPA measurements using squeezed light, as the QFI for squeezed states diverges in the limit of small absorbances. Focusing on particular measurement setups, we found that TPA absorption can be estimated with a CFI that shows super-Heisenberg scaling with the mean number of photons. In particular, for the case of very low absorbance, photon counting measurements for an input coherent state shows a scaling $\propto n^3$, which is even greater for quadrature measurements of a squeezed state, featuring a scaling $\propto n^4$. These scalings cannot be achieved in coherent second-harmonic generation, where the CFI scales quadratically for both coherent and squeezed states.

Future research should extend these results to the multimode regime, where in addition to the photon statistics considered in this work time-energy entanglement provides an additional resource, whose impact on TPA measurements is the subject of an intense current debate~\cite{Tabakaev2020, Raymer2020, Landes2020, Parzuchowski2021}.

\begin{acknowledgments} 

The authors thank Maria V. Chekhova and Manuel Gessner for helpful discussions. 
F. S. acknowledges support from the Cluster of Excellence 'Advanced Imaging of Matter' of the Deutsche Forschungsgemeinschaft (DFG) - EXC 2056 - project ID 390715994. 
C. S. M. acknowledges that the project that gave rise to these results received the support of a fellowship from la Caixa Foundation (ID 100010434) and from the European Union's Horizon 2020 Research and Innovation Programme under the Marie Sklodowska-Curie Grant Agreement No. 47648, with fellowship code  LCF/BQ/PI20/11760026.

\end{acknowledgments}

\bibliography{bibliography_photons}
\bibliographystyle{mybibstyle}

\onecolumngrid
\appendix

\newpage
\section{Squeezed and displaced bases}

\subsubsection{Squeezed vacuum}
The density matrix of a squeezed vacuum state is given by 
\begin{equation}
\rho_0 = S(\zeta) \vert 0 \rangle \langle 0 \vert S^\dagger (\zeta),
\end{equation}
where $\zeta = r e^{i \varphi}$ is the squeezing parameter, and $\vert 0 \rangle$ the vacuum state. It is usually convenient to evaluate correlation functions with respect to this operator in the Heisenberg picture, where the operators are transformed as \cite{BarnettRadmore}
\begin{equation}
a' = S^\dagger a S = \cosh (r) a + e^{i \varphi} \sinh(r) a^\dagger \label{eq:squeezing-transformation}
\end{equation} 
and 
\begin{equation}
a'^\dagger = S^\dagger a^\dagger S = \cosh (r) a^\dagger + e^{- i \varphi} \sinh(r) a.
\end{equation}

\subsubsection{Coherent state}
A coherent initial state is given by 
\begin{equation}
\rho_0 = U(\alpha) \vert 0 \rangle \langle 0 \vert U^\dagger (\alpha),
\end{equation}
and 
the Heisenberg evolution of the photon annihilation operator reads 
\begin{equation}
a' = U^\dagger a U =a + \alpha. \label{eq:displacement}
\end{equation}

\section{Quantum Fisher information}

Information on the TPA losses is encoded by letting the system evolve according to Eq.~(\ref{eq.Lindblad}) for a certain time,
\begin{equation}
\rho_{\varepsilon}=e^{\mathcal{\varepsilon}\mathcal{L}}\rho(0).
\end{equation}
The QFI $\mathcal{F}_{Q} (\rho)$ associated with the estimation of $\varepsilon$ can be obtained from the symmetric logarithmic derivative (SLD) $L_{\varepsilon}$ as $F_{Q}=\mathrm{Tr}[L_{\varepsilon}^{2}\rho_{\varepsilon}]$,
where the SLD is defined as to fulfil Eq.~(\ref{eq:SLD}).
Since $d\rho_{\varepsilon}/d\varepsilon=\mathcal{L}\rho_{\varepsilon}$,
we can transform Eq.~\eqref{eq:SLD} into
\begin{equation}
\frac{1}{2}\left( L_{\varepsilon}\rho_{\varepsilon}+\rho_{\varepsilon}L_{\varepsilon}\right)=\mathcal{L}\rho_{\varepsilon}. \label{eq.def-SLD}
\end{equation}
Taking matrix elements of both sides of this equation, we obtain Eq.~(\ref{eq:Lepsilon}). For squeezed vacuum or a coherent state, we can even solve this analytically. 

\subsubsection{Squeezed vacuum}

\begin{figure}[t]
\centering
\includegraphics[width=0.48\textwidth]{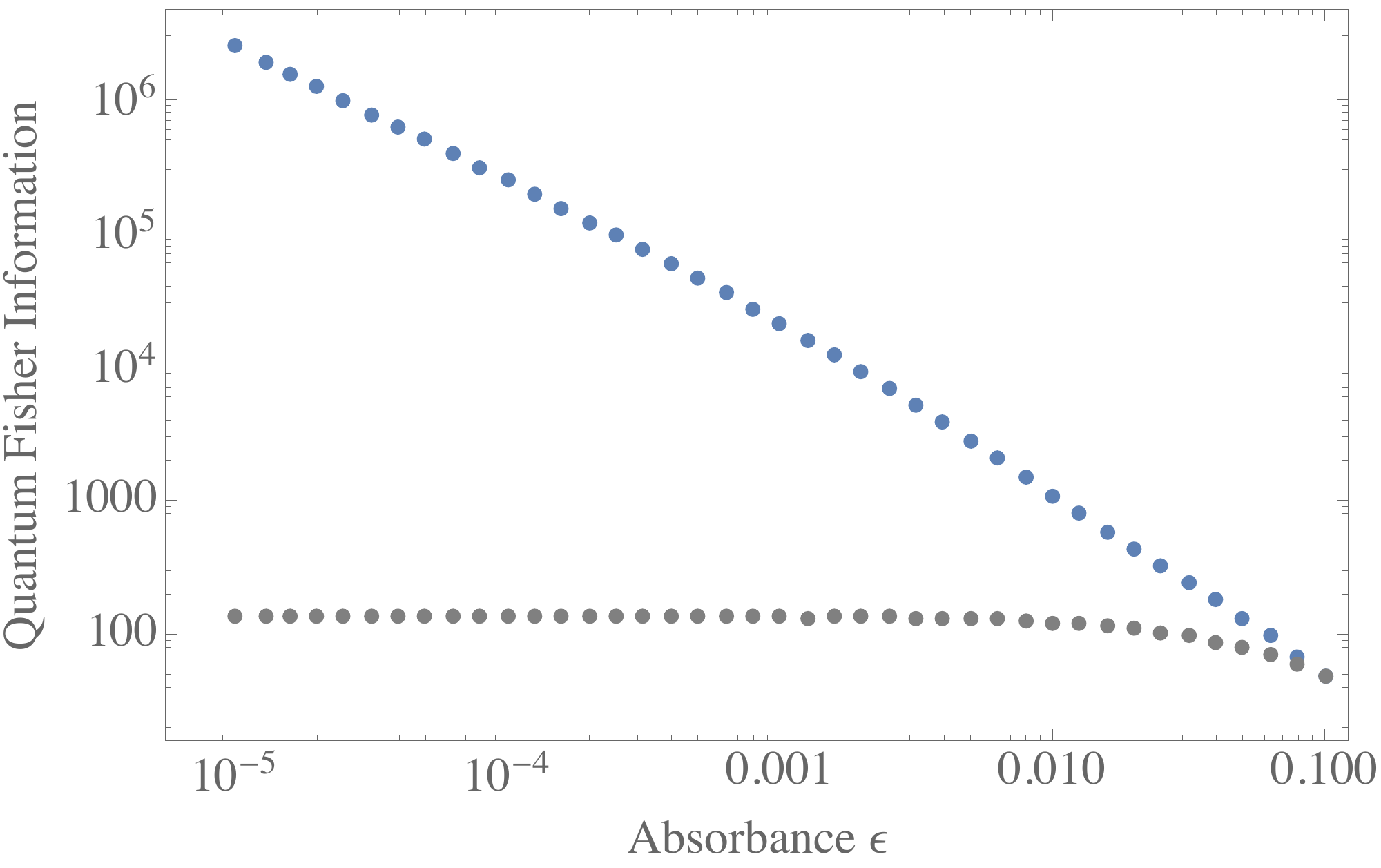}
\caption{(Color online) 
Quantum (QFI) and classical Fisher Information (CFI) vs. the TPA absorbance $\varepsilon = \gamma_{\mathrm{TPA}} t$. The QFI of a squeezed initial state (blue) and of a coherent state (red) are compared to the CFI for photon number measurements of the same squeezed (grey) and coherent state (black). The red and black dots coincide and cannot be distinguished.
}
\label{fig.FI-vs-epsilon}
\end{figure}

Let us consider the case $\varepsilon=0$, such that in the Heisenberg picture the initial state is simply $\rho_\varepsilon = \rho(0)=|0\rangle \langle 0|$. We can then analytically compute $L_0$ in
the squeezed basis, where we have:
\begin{equation}
\frac{1}{2}\left(L_{0}|0\rangle\langle0|+|0\rangle\langle0|L_{0}\right)=\mathcal{L}|0\rangle\langle0|.
\label{eq:SLD-00}
\end{equation}
By application of Eq.~\eqref{eq.Lindblad} with the transformed operators in Eq.~\eqref{eq:squeezing-transformation}, $\mathcal{L}|0\rangle \langle 0|$ reads
\begin{multline}
\mathcal{L}|0\rangle\langle0|=-4\sinh^{4}r|0\rangle\langle0|+4\sinh^{4}r|2\rangle\langle2|
+\left\{\frac{e^{-i\theta}}{\sqrt{2}}\left[\sinh(2r)-\sinh(4r)\right]|0\rangle\langle2|+\mathrm{h.c.}\right\}\\
-\left\{\frac{\sqrt{6}}{2}e^{-2i\theta}\sinh^{2}(2r)|0\rangle\langle4|+\mathrm{h.c.}\right\}\\
\label{eq:Lro0}
\end{multline}
Here, we see that the application of the TPA Liouvillian on the vacuum in the squeezed basis creates a population $|2\rangle\langle2|$. This matrix element cannot be obtained by the single application of a bounded operator $L_0$. As a consequence, the resulting SLD operators are divergent. We can verify this numerically in Fig.~\ref{fig.FI-vs-epsilon}, where the QFI of squeezed states diverges for $\varepsilon \rightarrow 0$.

Transforming back to the normal basis, this matrix element turns into the population of a squeezed two-photon Fock state, $S(\zeta) |2\rangle\langle2| S^\dagger (\zeta)$. As a consequence, a projective measurement with $P = S(\zeta) |2\rangle\langle2| S^\dagger (\zeta)$ will be an optimal measurement to determine TPA losses of a squeezed vacuum state. 
We note in passing that the same also applies for incoherent single photon losses (i.e. when in Eq.~(\ref{eq.Lindblad}) the Lindblad operator is $L = a$), where a similar analysis as above will reveal the population of the state $S(\zeta) |1\rangle\langle1| S^\dagger (\zeta)$, which again represents an optimal measurement. This divergence takes place, even though single photon losses never create non-Gaussianity in the time-evolved state. The Wigner function of the evolved state will always remain positive.

\subsubsection{Coherent light}

As before, we compare matrix elements of the constituent equation in the displaced basis,
\begin{equation}
\frac{1}{2}\left(L_{0}|0\rangle\langle0|+|0\rangle\langle0|L_{0}\right)=\mathcal{L}|0\rangle\langle0|.
\label{eq:SLD-00}
\end{equation}
We find 
\begin{equation}
\mathcal{L}|0\rangle\langle0| = - |\alpha|^2 \alpha |1\rangle\langle0| - \frac{\alpha^2}{\sqrt{2}} |2\rangle\langle0| + h.c.
\end{equation}
The Liouvillian in the displaced basis does not create nonzero populations, as the displacement does not mix creation and annihilation operators. Consequently, we can identify the nonzero matrix elements of the SLD operators, 
\begin{align}
L_{00} &= 0\\
L_{10} &= L_{01}^* = - |\alpha|^2 \alpha\\
L_{20} &=L_{02}^*= \frac{1}{\sqrt{2}} \alpha^{2}\\
L_{03} &=L_{30}=0\\
L_{04} &= L_{40}= 0
\end{align}
Using the SLD operator, we straightforwardly calculate the QFI of a coherent state at $\varepsilon = 0$,
\begin{equation}
\mathcal{F}_{\rho_{\mathrm{coh}}} ( \varepsilon = 0 ) = \lim_{\varepsilon\rightarrow 0 }  \mathrm{Tr}[L_\varepsilon^2 \rho_\varepsilon] = \sum_{i=0}^4 |L_{0i}|^2 \\
= n_\alpha^3 + \frac{1}{2} n_\alpha^2.
\end{equation}

\section{Classical Fisher information}

The classical Fisher information is defined in Eq.~(\ref{eq.F_C}). We now evaluate it for different measurements. 

\subsection{Photon number measurements}
Here, the probabilities are given by $p_n = \langle n \vert \rho \vert n \rangle$. To evaluate this, we note that the action of the TPA Lindbladian~(\ref{eq.Lindblad}) on a density matrix element $\vert n \rangle \langle m \vert$ yields
\begin{align}
\mathcal{L} \vert n \rangle \langle m \vert &= \frac{1}{4} \left( 2 \sqrt{n(n-1)m(m-1)} \vert n-2 \rangle \langle m-2 \vert - \left( n (n -1) + m (m-1) \right)  \vert n \rangle \langle m \vert \right).
\end{align}
Consequently, the change of photon number distribution due to TPA is given by
\begin{align}
\frac{ d P_n }{ d\varepsilon } &= \frac{1 }{2} \left( (n+2) (n+1) P_{n+2} - n(n-1) P_{n} \right),
\end{align}
where $p_{n,0}$ denotes the probability to detect $n$ photons prior to the interaction with the TPA medium, and we arrive at
\begin{align}
\mathcal{F}_C (\rho, \hat{n}) &= \sum_n \frac{1}{4} \frac{ (  (n+2) (n+1) P_{n+2}  - n(n-1) P_{n} )^2 }{P_{n}}. 
\end{align}

\subsection{Quadrature measurements}

We want to construct the classical Fisher information related to quadrature measurements, i.e. for the measurement based on a continuous probability distribution $P (q)$ (or $P (p)$).
For the $q$-quadrature, it reads, for instance,
\begin{equation}
\mathcal{F}_C (\rho_\varepsilon, q) = \int dq \frac{1}{P (q)}\left(\frac{ dP(q)}{d\varepsilon }\right)^2. \label{eq.quad-CFI}
\end{equation}
The necessary probability distributions can be constructed most conveniently from the Wigner functions of the photonic quantum states using the relation~\cite{Brune1992}
\begin{equation}
W (\alpha) = \frac{2}{\pi} e^{- 2 |\alpha|^2} \Re \left[ \sum_{n\geq m} (-1)^m (1 - \delta_{n m}) \left( \frac{m!}{n!} \right)^{n-m} (2\alpha)^{n-m} \mathcal{L}_m^{n-m} (4 |\alpha|^2) \rho_{mn} \right], \label{eq.W(q,p)}
\end{equation}
where $\alpha = q + i p$, $\mathcal{L}_m^{n-m}$ is a Laguerrre polynomial and $\rho_{mn}$ the density matrix element in the photon number basis.
Integrating out the conjugate variable, we obtain the probability distribution, i.e.
\begin{equation}
P (q) = \int \!\! dp \; W (q, p).
\end{equation}
Similarly, we can calculate the change of the Wigner functions due to TPA losses,
\begin{equation}
\frac{dW (q, p)}{d\varepsilon} = \frac{2}{\pi} e^{- 2 |\alpha|^2} \Re \left[ \sum_{n\geq m} (-1)^m (1 - \delta_{n m}) \left( \frac{m!}{n!} \right)^{n-m} (2\alpha)^{n-m} \mathcal{L}_m^{n-m} (4 |\alpha|^2) (\mathcal{L}\rho)_{mn} \right],
\end{equation}
and use it to straightforwardly calculate the change of the probability distribution
\begin{equation}
\frac{dP (q)}{d\varepsilon} = \int_{-\infty}^{\infty} \!\!\!\! dp \; \frac{dW (q, p)}{d\varepsilon}.
\end{equation}

\subsubsection{Squeezed vacuum}

For a squeezed vacuum state, we use the transformation~(\ref{eq:squeezing-transformation}) to carry out the above analysis in the squeezed vacuum basis, where the only nonzero matrix element in Eq.~(\ref{eq.W(q,p)}) is the vacuum state $\rho_{00}$. Hence, we have 
\begin{equation}
P_{squ} (q) = \sqrt{\frac{2}{\pi}} e^{r - 2 e^{2r} x^2}
\end{equation}
and
\begin{equation}
\frac{dP_{squ} (q)}{d\varepsilon} = - \frac{1}{\sqrt{2\pi}} e^{r - 2 e^{2r} x^2} \sinh (r) \left( e^{2r} + 6 x^2 + 2 e^{3r} x^2 \left( \sinh (r) - 5 \cosh (r) + 8 x^2 \sinh (r) \right) \right).
\end{equation}
With Eq.~(\ref{eq.quad-CFI}), we thus arrive at
\begin{equation}
\mathcal{F}_C (\rho_{squ}, q) = \frac{e^{-2r} \sinh^2(r)}{8} \left( 4 e^{8r} - 12 e^{6r} +33 e^{4r} - 42 e^{2r} +21 \right).
\end{equation}
For large squeezing, the photon number is approximately given by $n_r \simeq e^{2r} /4$, such that we arrive at Eq.~(\ref{eq.F_C(X)}) of the main text. 

An identical calculation yields the CFI for the $p$-quadrature
\begin{equation}
\mathcal{F}_C (\rho_{squ}, p) = \frac{e^{-6r} \sinh^2(r)}{8} \left( 21 e^{8r} - 42 e^{6 r} + 33 e^{4r} - 12 e^{2r} + 4 \right).
\end{equation}

\subsubsection{Coherent state}

Using Eq.~(\ref{eq:displacement}), the same calculation as for the squeezed vacuum above yields
\begin{equation}
\mathcal{F}_C (\rho_{\mathrm{coh}}, q) = n_{\alpha}^3 + \frac{n_{\alpha}^2}{2}
\end{equation}
and 
\begin{equation}
\mathcal{F}_C (\rho_{\mathrm{coh}}, p) = \frac{n_{\alpha}^2}{2}.
\end{equation}

\subsubsection{Classical Fisher information vs. absorbance}
The absolute value of the classical Fisher Information which are used to extract the scaling properties shown in Fig.~\ref{fig.phase-space} of the main text is shown in Fig.~\ref{fig.FCI-phase-space} below.

\begin{figure}[b]
\centering
\includegraphics[width=0.8\textwidth]{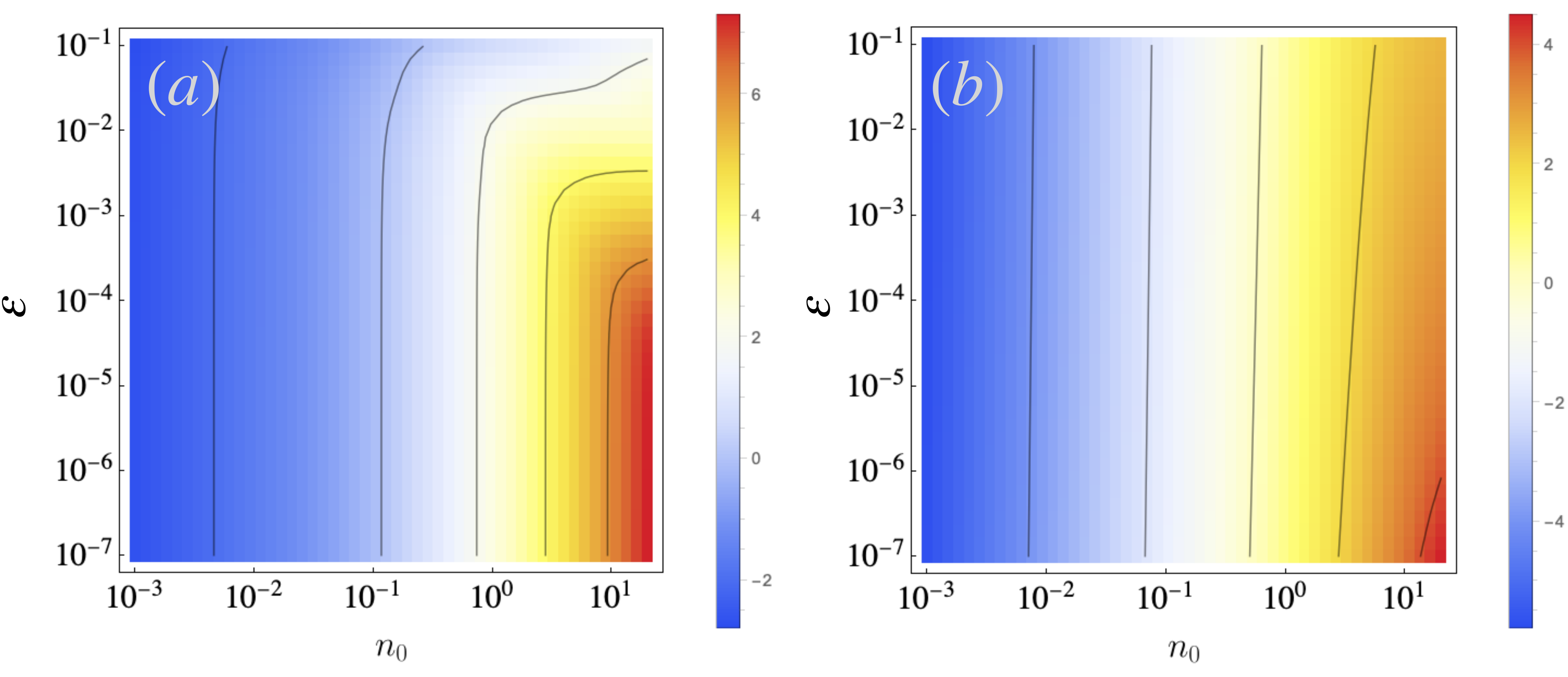}
\caption{(Color online) 
(a) The classical Fisher information~(\ref{eq.F_C}) for quadrature measurements of the squeezed quadrature is plotted vs. the photon number $n_0$ and absorbance $\varepsilon$ of a squeezed vacuum state.
(b) The same as in (a) for a coherent input state. 
}
\label{fig.FCI-phase-space}
\end{figure}

\end{document}